\documentclass[12pt]{article}
\usepackage{a4wide}
\usepackage{amssymb}
\usepackage{graphicx}
\begin{document}
{\renewcommand{\thefootnote}{\fnsymbol{footnote}}
\medskip
\begin{center}
{\LARGE Quantum cosmology: a review}\\
\vspace{1.5em}
Martin Bojowald\footnote{e-mail address: {\tt bojowald@gravity.psu.edu}}
\\
\vspace{0.5em}
Institute for Gravitation and the Cosmos,\\
The Pennsylvania State
University,\\
104 Davey Lab, University Park, PA 16802, USA\\
\vspace{1.5em}
\end{center}
}

\setcounter{footnote}{0}

\begin{abstract}
  In quantum cosmology, one applies quantum physics to the whole
  universe. While no unique version and no completely well-defined theory is
  available yet, the framework gives rise to interesting conceptual,
  mathematical and physical questions. This review presents quantum cosmology
  in a new picture that tries to incorporate the importance of inhomogeneity:
  De-emphasizing the traditional minisuperspace view, the dynamics is rather
  formulated in terms of the interplay of many interacting ``microscopic''
  degrees of freedom that describe the space-time geometry. There is thus a
  close relationship with more-established systems in condensed-matter and
  particle physics even while the large set of space-time symmetries (general
  covariance) requires some adaptations and new developments. These extensions
  of standard methods are needed both at the fundamental level and at the
  stage of evaluating the theory by effective descriptions.
\end{abstract}

\newpage

\tableofcontents

\section{Introduction}

Quantum cosmology is based on the idea that quantum physics should apply to
anything in nature, including the whole universe.\footnote{The definition of
  ``quantum cosmology'' has changed over the years from quantization
  restricted to the degrees of freedom of highly symmetric classical
  cosmological spacetimes to the broader viewpoint taken here.}  Quantum
descriptions of all kinds of matter fields and their interactions are well
known and can easily be combined into one theory --- leaving aside the more
complicated question of unification, which asks for a {\em unique} combination
of all fields based on some fundamental principles or
symmetries. Nevertheless, quantizing the whole universe is far from being
straightforward because, according to general relativity, not just matter but
also space and time are physical objects. They are subject to dynamical laws
and have excitations (gravitational waves) that interact with each other and
with matter. Quantum cosmology is therefore closely related to quantum
gravity, the quantum theory of the gravitational force and space-time. Since
quantum gravity remains unfinished, the theoretical basis of quantum cosmology
is unclear. And to make things worse, there are several difficult conceptual
problems to be overcome.

For a theory that stubbornly retains its highly speculative and controversial
nature, quantum cosmology has a rather long history. Soon after the basic
ingredients of general relativity and quantum mechanics had become known,
adventurous theorists began to apply these new-found laws to the
cosmos. Lema\^{\i}tre's ``primordial atom''
\cite{LemaitreCosmo,LemaitreNature} combines insights from general relativity
(what would later be called the big-bang singularity) with key concepts of
quantum mechanics. Tolman's attempts \cite{TolmanEntropy,Tolman} to use
quantum physics in order to solve the singularity issue of general relativity
were in many ways prescient and brought up problems still relevant today,
especially the issue of entropy growth in an eternal universe. But further
progress was hindered not only by the incomplete status of quantum mechanics
at the time or a lack of interest from mainstream theorists. (After all,
atomic physics provided much more practical and pressing problems.) More
importantly, the mathematical issues to be solved when one tried to go beyond
the simplest models were daunting. Even today, more than 80 years later, the
setting remains incompletely realized in several different, sometimes
competing ways in which one might be able to quantize the gravitational force:
string theory, canonical quantum gravity, and non-commutative geometry to name
the most popular ones. And when it comes to concrete early-universe scenarios
``derived'' from any one of these frameworks, the number of different and
often mutually contradictory attempts rapidly multiplies.

Moreover, unlike other areas of physics that were largely reformed or newly
created with the arrival of quantum mechanics, the application of quantum
physics to the whole universe has not been able to make contact with
observations, nor has it been able to resolve conceptual issues to a degree
which would allow one to say that it works for all practical purposes. Then
why would anyone be interested in quantum cosmology? In spite of all
misgivings, quantum cosmology offers not only challenges but also interesting
insights about a variety of questions, most importantly the fundamental nature
of space-time. There are also stimulating mathematical and conceptual
questions, related for instance to the interplay of symmetries and discrete
structures or the extraction of semiclassical features from a generally
covariant and highly interacting quantum theory. This review attempts to give
a modern viewpoint that takes into account previous roadblocks and points out
promising activities.

\subsection{The premise of quantum cosmology}

The main part of this review is the presentation of quantum cosmology based on
a physical and microscopic picture of quantum space-time, somewhat similar to
atomic systems studied in condensed-matter physics.  It turns out that this
condensed-matter analogy may not only serve as a pedagogical tool to make a
somewhat exotic field more broadly accessible; it will also allow one to
transfer useful methods and potentially lead to cross-fertilization between
different fields. One could, moreover, hope that the down-to-earth nature of
systems studied in condensed-matter physics will lead to a more humble view on
quantum cosmology, which tends to be attracted to grandiose questions
regarding the whole universe, everything in or about it, and the origin.

We first set up the problems to be solved by quantum cosmology. General
relativity has shown that space-time is physical and dynamical, and that it
interacts with matter. Since matter is described by quantum physics, a
consistent coupling requires space-time to be quantized as well. In the
absence of any complete theory of quantum gravity (let alone experiments),
however, the structure of microscopic degrees of freedom of space-time remains
unknown. Even the classical nature of space-time, which will likely play the
role of the continuum limit of some more fundamental quantum space-time, had
not been uncovered until special relativity was developed. As in this
well-known case, conceptual arguments play an important role in the approach
to the quantum nature of space-time, providing guidelines in the absence of
direct observations.

As shown by special relativity, and even more so by general relativity,
space-time is characterized by the symmetries it enjoys. These symmetries of
general covariance are not accidental but ensure that the theory is
meaningful, in the sense that predictions do not depend on the choice of
observers or mathematical descriptions and coordinates.  They are not to be
broken by quantum effects or else the theory will be inconsistent. Unlike,
say, crystals that may break the rotational symmetry of the vacuum, space-time
must be treated much more delicately if it is to be described in microscopic
terms. This problem is a difficult one: the various approaches to quantum
gravity and quantum cosmology are still looking for a consistent treatment.
But recent progress has shed some light on possible outcomes.

After introductory discussions in the present section, in Sec.~\ref{s:Single}
we enter details provided by different approaches. We begin with the analog of
one-particle Hamiltonians in order to understand the building blocks we are
dealing with. Just as quantum mechanics provides the laws for a quantum
description of idealized point-like particles, quantum cosmology should start
with a quantization of what might be considered the most elementary form of
space: a single point or a small uniform region around it. Clearly, without
direct observations, it is difficult to guess what the elementary constituents
should be. But the strong and difficult consistency conditions imposed by the
symmetries of space-time provide an advantage: they can fully be implemented
only by tightly controlled conceptual reasoning, which in many cases is so
restrictive that fundamental properties can be inferred theoretically. As
always, of course, theoretical derivations are based on some principles which
may very well be wrong as far as Nature is concerned. Nevertheless, consistent
sets of suitable laws for different forms of classical or quantum space-time
are rare, a feature which, once consistency has been fully implemented, should
strongly reduce the arbitrariness in possible forms of constituent laws.

\subsection{Problems to be faced}

Quantum cosmology starts with a quantization of a structureless, homogeneous
chunk of space as a first approximation to a ``space-time atom.''
Traditionally \cite{DeWitt}, the resulting systems have, under the name of
minisuperspace models, been thought of as quantizations of a whole, spatially
homogeneous universe that might approximate the actual one quite well at least
at early times. However, to anticipate a little bit, there are several
problems\footnote{This review discusses a number of key problems, introduced
  in due course. A problem in this class does not necessarily indicate a
  failure of the approach in which it appears, but it highlights important
  issues to be understood and therefore helps to focus research. Some problems
  may be considered more or less severe by different researchers. As always,
  the viewpoints taken in this review are partially affected by the author's
  opinion and degree of familiarity with other approaches, but an attempt has
  been made to focus on problems that are widely recognized in the general
  field of quantum gravity and cosmology (but not always in a single
  approach).}  with this picture, not just because the sense in which an
actual approximation is achieved is rather uncontrolled. For this reason, we
take what one may think of as the opposite view, understanding these systems
not as models for the whole universe but rather as quantizations of the
smallest possible and structureless contributions to space.

Another conceptual problem can already be glimpsed at this stage: Quantum
cosmology attempts to find the correct quantum theory of {\em space-time}. But
our brief sketch of a quantum-mechanical model of the elementary constituents
states that we should quantize a point (or small region) in {\em space}. If we
were to follow standard quantum mechanics, time would be provided as an
external, un-quantized parameter, playing a role very different from the
quantized chunk of space to be described. Such a theory could hardly be
covariant. The nature of time, and not just the covariance symmetries it is
crucially involved in, must therefore be rethought for a fundamental theory of
quantum cosmology.

Given a consistent quantum model of building blocks of space (perhaps evolving
in time), a general example of quantum space is then a collection of these
chunks, patched together to form a large system which over long distances must
look like the curved continuum of general relativity. The collection of chunks
of space may be viewed as an approximation to inhomogeneous space by simpler
homogeneous pieces as it may be of some use for a classical evaluation. But in
a more fundamental view, it could also correspond to an actual physical
decomposition of space if quantum gravity leads to a discrete spatial
structure, as some approaches do. Two tasks are then to be faced: (i) Find the
correct quantum theory for a single patch, or the single-patch theory
analogous to one-particle dynamics, and (ii) determine the rules to combine
patches into a many-patch system, analogous to interacting many-body
systems. When these tasks have been completed in a consistent form, quantum
cosmology has been constructed. It then remains the task of finding manageable
ways of evaluating the theory and comparing its predictions with possible
observations.

The derivation of single-patch theories starts with the classical dynamics,
given by the Friedmann equation of cosmology. At this point, the somewhat
problematic role of time can already be discussed because the system is
reparameterization invariant (the time coordinate can be changed at will),
giving a first look on issues involved with covariance at a quantum
level. The main approach used here is canonical quantum cosmology, which
directly quantizes the Friedmann equation. (However, other approaches could be
interpreted as taking comparable, although technically quite different,
single-patch views, for instance string theory with a single string worldsheet
as a test particle.) A canonical quantization of the Friedmann equation will
therefore present candidates for single-patch theories. We emphasize again
that the single-patch viewpoint does not amount to a traditional
minisuperspace quantization, although they both quantize the Friedmann
equation and are formally equivalent in some approaches. In a minisuperspace
model, one typically interprets the patch as the whole universe, approximately
homogeneous early on before structures had time to form. This traditional
viewpoint has led to several problems, especially in discrete theories which
give rise to quantum corrections that depend on the discrete patch size. In a
minisuperspace model, the magnitude of these corrections is not captured
properly because one is led to insert the whole size of the visible universe
for the microscopic patch size.

Moreover, owing to the complicated symmetry structure in a covariant theory,
there is a big difference between an exactly homogeneous model and one
slightly perturbed by inhomogeneous structure. Even if classical back-reaction
of inhomogeneous modes on the homogeneous background can be ignored,
inhomogeneity leads to much tighter constraints on the dynamics, even of the
background, than an exactly homogeneous minisuperspace model would suggest. It
is much more difficult to ensure covariance of (inhomogeneous) partial
differential equations compared to (exactly homogeneous) ordinary
ones. Compared with the symmetry of matter distributions, which as an
approximation often simplifies but still approximates realistic physical
systems, space-time covariance is more subtle. It removes redundant degrees of
freedom such as choices of coordinates, not directions along which structure
may, at leading order, be unresolved.  For these reasons, in this review we
try to avoid the somewhat misleading minisuperspace view, which makes use of
large exactly homogeneous spaces, and instead refer to single
patches. (Minisuperspace models are discussed briefly in a historical
context.)

After a detailed discussion of single-patch theories, qualitative patching
models are described in Sec.~\ref{s:Patching}. There are two main examples,
the first one derived from the classical Belinski--Khalatnikov--Lifshitz
scenario \cite{BKL}, according to which even highly inhomogeneous space
behaves near a singularity, such as the big bang, like a collection of
disconnected homogeneous pieces. This scenario leads to a patch model in which
constituents are non-interacting initially (near the singularity), but slowly
build up more complicated dynamics as the universe expands and dilutes to
smaller densities. The second example of qualitative patching models is more
recent and uses well-studied techniques from condensed-matter physics, in
particular mathematical descriptions of Bose--Einstein condensation. Different
patches of space (unlike individual particles in a condensate) cannot be
required to occupy exactly the same state, but in a nearly homogeneous
universe all the individual states should at least be close to one
another. Mathematical methods of Bose--Einstein condensation then map the
interacting many-patch dynamics to a non-linear single-patch dynamics of
states, described by some version of the Gross--Pitaevski equation.

The exact patching of single contributions in all its details is supposed to
be determined by quantum gravity, which remains incomplete. Nevertheless,
several interesting indications already exist and can be exploited in models
of quantum cosmology. Section~\ref{s:Many} is devoted to these
consequences, but it is brief.

Any potential candidate for quantum cosmology or quantum gravity must be
evaluated and scrutinized.  Another main section of this review,
Sec.~\ref{s:Eff}, introduces and deals with effective theories as the
method of choice. Here, again, there is some overlap with
condensed-matter techniques, at least at a technical level. Many-patch
dynamics is usually too complicated to be dealt with directly, and in quantum
gravity even numerical methods are still in their infancy and do not yet help
much. Low-energy and semiclassical methods are therefore essential, by which
one can approximate complicated quantum theories in interesting and
observationally accessible regimes. There is a large set of techniques from
different approaches, such as double-sheeted universe models (not unlike
double layers of graphene) or deformed coordinate structures in
non-commutative geometry, brane-world models in string theory, and different
types of effective equations in canonical quantum cosmology.

Canonical effective methods are perhaps the most systematic. They begin with
the well-known Ehrenfest equations of quantum mechanics but can be amended to
take into account new structures required for cosmology, including a more
complicated dynamics and symmetry issues such as covariance and independence
of one's choice of time. The important and recurring problem of time in
quantum gravity is discussed more fully in this context.

Also the problem of states plays a role: Most effective techniques make use of
a simple ground, vacuum or other state around which one can expand, a
reasonable choice for near-stationary and lowly-excited properties of systems
in condensed-matter or particle physics.  But if one tries to extend these
methods to quantum cosmology, one encounters the problem that it remains
unclear what the vacuum of quantum space should be, a state in which,
presumably, no space-time and no geometry are excited. Moreover, even if a
vacuum state existed, one would need high excitations of a large number of
patches to deal with macroscopic, even cosmic and potentially high-density
regions of space. One must find suitable many-body states, or perhaps an
analog of finite-temperature states in which not the distribution of particle
velocities but the arrangement of spatial patches is close to the classical
continuum. Such states are difficult to construct and remain unknown, but some
of their properties can be implemented by effective methods, also shedding
light on the form of quantum space-time symmetries.

\subsection{Opportunities}

The challenges encountered by quantum cosmology provide us with an opportunity
to revisit general methods to construct and evaluate physical quantum
systems. Especially the theory of effective descriptions, but also general
quantization procedures, must be extended considerably if one tries to
address the problems of time, covariance, and states. In such an analysis, one
may find new insights that could be useful also in other systems. Certainly
from a mathematical perspective there are already several interesting
examples.

A notion that should be expected in the context of many-patch systems,
considered at vastly different scales that bridge fundamental candidates with
potential observations, is renormalization. It is not discussed in this
review because its formulation for quantum cosmology is still to be
attempted. Quantum space-time does not provide an energy scale by which one
could control the renormalization flow, and therefore even the basic set-up
remains unclear. (See \cite{Nets} for a model of possible implementations.)
But the question is certainly important and presents an interesting challenge.

Models of quantum cosmology help to derive and understand possible candidates
for the quantum structure of space-time. This issue plays an important role
for quantum gravity as well, but is difficult to analyze at such a general
level that includes all possible solutions. The restriction to cosmological
space-times with some approximate symmetries, in addition to the potential for
observations at high density, provides additional tools and motivations.

Physically, the ultimate pay-off of theoretical developments is the
confrontation with experimental data.  The final section of this review gives
an outlook on potential observational tests of the theory. In the context of
quantum gravity one usually thinks of the Planck density $\rho_{\rm
  P}=c^7/\hbar G^2$, formed by the fundamental constants of gravity,
relativity and quantum physics, which is much larger by many orders of
magnitude than the average density encountered in any observationally
accessible regime. (It amounts to more than a trillion solar masses in a
region the size of a single proton. Similarly, the distance measure given by
the Planck length $\ell_{\rm P}=\sqrt{G\hbar/c^3}\approx 10^{-35}{\rm m}$ is
much tinier than anything that can be probed by particle accelerators.)
Moreover, while quantum gravity remains incomplete and usually subject to a
large number of quantization ambiguities, predictions in the Planck regime are
uncontrolled. But indirect effects at lower density --- in which small
individual quantum corrections of many patches taken together conspire to form
a more sizeable contribution --- are possible, much like Brownian motion which
helped to detect atomic and molecular properties. There are two main examples:
large extra dimensions as postulated by string theory (which have been
reviewed extensively elsewhere), and different types of modifications from
discrete theories such as loop quantum gravity.

\section{Single-patch theories}
\label{s:Single}

In quantum cosmology, the analog of a single particle as an ingredient of
more-complicated interacting theories is a small and uniform chunk of
space. It is subject to simple laws in general relativity, which one can try
to quantize by familiar methods. The result, which is formally equivalent to
traditional minisuperspace models of quantum cosmology, is called here a
single-patch theory. In several conceptual and practical aspects the
single-patch viewpoint differs crucially from a minisuperspace picture. It
turns out to be better suited for going beyond the highly simplified setting
of exactly homogeneous cosmology in a realistic description of the evolution
of structure in the universe.

In spite of such advantages, it is far from clear whether a single-patch
theory (or a minisuperspace model) captures the correct fundamental degrees of
freedom of quantum space-time. After all, a water molecule is not a water
droplet. The whole process of quantizing general relativity or constructing
quantum gravity has occasionally been questioned based on an analogy with
hydrodynamics, in which applying quantum rules to the Navier--Stokes equation
would not give the right fundamental theory (see for instance
\cite{QuantumHydro,QBR}). Only observations can, in this case, show what the
fundamental constituents of water are. Trying a quantization strategy on
space-time without any observations of fundamental constituents may, from this
perspective, seem preposterous.

However, there is one indication why an attempted quantization of a chunk of
space should not be meaningless, although it would certainly be highly
simplified compared to general space-time configurations. A chunk of space
evolving in time can be seen as a local version of general
relativity. Locally, the most important and basic space-time structure of this
classical theory is given by the form of Lorentz or Poincar\'e symmetries. The
same symmetries are known to play a fundamental role in high-energy physics,
where they determine mathematical representations of elementary particles and
other properties. The intimate relation to fundamental symmetries makes a
chunk of space play a more reliable role for quantum gravity than a water
droplet does for molecular physics of liquids. There is no guarantee that this
strategy will lead to the right physical result, which ultimately can be
tested only by experiments. But as long as the theoretical infrastructure
needed to start probing the theory observationally is still being developed,
the approach to quantum cosmology sketched here is promising.

\subsection{Classical dynamics}

The geometrical property of isotropic space, which becomes physical and
dynamical in general relativity, is its volume. There are two ways in which a
volume parameter can be assigned to a given region ${\cal V}$ in space. First,
assuming some set of coordinates $x^a$, $a=1,2,3$, covering the region, we can
simply integrate and define the coordinate volume $V_0=\int_{\cal V}{\rm
  d}^3x$. However, the result depends not just on the region but also on the
choice of coordinates. To make the volume coordinate independent, we use a
metric tensor $h_{ab}$ and define $V=\int_{\cal V}\sqrt{\det h}\: {\rm
  d}^3x$. The result now depends on the choice of metric rather than
coordinates. But the metric is considered a physical field in general
relativity, and the volume is a measurable quantity by which properties of the
metric can be determined.

For a homogeneous, isotropic and flat chunk of space, the metric
$h_{ab}=a(t)^2\delta_{ab}$ in Cartesian coordinates has only one free
function, the scale factor $a(t)$ depending on time. General relativity
requires that this function obeys the Friedmann equation
\begin{equation} \label{Friedmann}
 \left(\frac{\dot{a}}{a}\right)^2= \frac{8\pi G}{3c^2} \rho
\end{equation}
with the energy density $\rho$ of matter.  (On the left-hand side, there could
be additional geometry contributions depending on the scale factor, $kc^2/a^2$
with $k=\pm 1$ if space has constant positive or negative curvature, and
$-\Lambda$ if there is a non-zero cosmological constant $\Lambda$.)

In order to derive the Friedmann equation, one extends the spatial metric to a
space-time metric $g_{\mu\nu}$ and inserts it in the general Einstein
equation. The extension, still respecting isotropy, can be done with a second
free function $N(t)$, so that, written as a line element,
\begin{equation}\label{Line}
 {\rm d}s^2=g_{\mu\nu}{\rm d}x^{\mu} {\rm d}x^{\nu}=
 -N(t)^2{\rm d}t^2+ a(t)^2 ({\rm d}x^2+{\rm
  d}y^2+{\rm d}z^2)\,,
\end{equation}
summing over four values of $\mu$ and $\nu$. Unlike $a(t)$, $N(t)$ has no
relation whatsoever with observations; it can simply be eliminated by
introducing a new (proper) time coordinate $\tau$, integrating ${\rm
  d}\tau=\int N(t){\rm d}t$. In the Friedmann equation (\ref{Friedmann}), we
have already assumed that the time derivative is taken with respect to proper
time; otherwise the fraction would be $\dot{a}/(Na)$. The whole equation is
then independent of the choice of time coordinate. In this way, the
time-reparameterization part of general covariance survives even in this
highly restricted setting of homogeneous cosmology.

With its energy contribution and a ``kinetic'' term quadratic in $\dot{a}$,
the Friedmann equation looks similar to an energy-balance law. However, there
are several differences to the standard form, which turn out to foreshadow the
problems to be faced by quantum cosmology. We can see this more clearly if we
rewrite the equation as
\begin{equation}\label{Friedmann2}
 -\frac{3c^2}{8\pi G}a\dot{a}^2+ a^3\rho=0\,.
\end{equation}
The matter energy density $\rho$ now appears in the combination $a^3\rho$,
which is the total energy $\int_{\cal V} \rho\sqrt{\det h}\: {\rm d}^3x$
contained in a region of coordinate volume $V_0=1$. Space-time geometry
contributes a term $-(3c^2/8\pi G)a\dot{a}^2$ with the right units of energy,
but one that is negative. A direct application of quantum mechanics with such
a Hamiltonian would not result in a stable ground state, implying the state
problem of quantum cosmology.

Another difference to classical mechanics is the fact that (\ref{Friedmann2})
does not define the energy as a free parameter or an observable, but instead
imposes a constraint because the sum of space-time and matter contributions
must always vanish. In quantum mechanics, with such a Hamiltonian one would
not obtain an evolution equation: the $i\hbar\partial/\partial t$-part
of the Schr\"odinger equation would be missing. Classically, this feature
implies that one can freely choose the time coordinate in which the equation
is written and solved. In quantum cosmology, the missing evolution picture
constitutes the problem of time.

The third problem, the one of covariance, is more difficult to see at this
stage because it shows its full force only when inhomogeneity is
included. Cartesian coordinates $x$, $y$ and $z$ are special and adapted to
the symmetry. (They are convenient but not required: in a generally covariant
theory, one may perform any transformation, not just the Euclidean ones that
would preserve the form of the metric.) A transformed metric
$h_{ab}\not=a(t)^2\delta_{ab}$, with position-dependent components, would not
look homogeneous, even though it would enjoy the same translational and
rotational symmetries and would predict the same physics. For covariance of
any quantization of the theory, one needs to make sure that a quantization of
the more complicated inhomogeneous-looking metric still provides the same
results. One can expect that quantum corrections must have a very specific
form for the function $a(t)$ solving a quantization (or an effective equation
with quantum corrections) of the simple isotropic (\ref{Friedmann2}) to be
part of a proper space-time metric that fulfills the covariance condition. We
will come back to these important but somewhat technical issues at a later
stage, and for now continue with exactly homogeneous models and their symmetry
of time reparameterization invariance.

Reparameterization invariance implies a symmetry with a generator that should
generalize the familiar energy functional, for time translation --- generated
by the energy --- is a simple version of time reparameterization. Also this
generator is provided by general relativity, most easily when one uses its
canonical formulation. (See \cite{CUP} for an introduction to the latter.) In
the cosmological setting, by taking derivatives of the action by a time
derivative of the degree of freedom $a$, we first obtain the momentum
$p_a=-(3c^2/4\pi G)V_0 a\dot{a}$ conjugate to the scale factor, and the
reparameterization generator
\begin{equation} \label{Friedmann3}
 C=-\frac{2\pi G}{3c^2}\frac{p_a^2}{V_0a}+V_0 a^3\rho 
\end{equation}
(which is (\ref{Friedmann2}) times $V_0$, expressed using $p_a$ instead of
$\dot{a}$).  The Friedmann equation amounts to the constraint equation
$C=0$. The zero, instead of an observable energy, indicates that
reparameterization invariance is not a symmetry that relates different viable
solutions but a gauge transformation which eliminates unphysical degrees of
freedom, such as any possible effect of the choice of time coordinates. In
whatever way one quantizes the theory, this gauge invariance must be
respected: There must be an operator $\hat{C}$ with $C$ as its classical
limit, so that physical states $\psi$ obey $\hat{C}\psi=0$.\footnote{This
  prescription is the so-called Dirac approach to quantizing
  constraints. Alternatively, one may solve $C=0$ completely, factor out the
  classical gauge flow, and then quantize the resulting reduced phase
  space. Owing to non-trivial phase-space structures, it is often more
  complicated to perform this procedure. Its results are not necessarily
  equivalent to those of the Dirac approach because the quantum gauge flow
  implemented in the latter procedure may differ from the classical gauge
  flow. Here, we focus on the Dirac approach in order to quantize the full
  gauge system.}  The classical invariance is then not destroyed, but there
may still be interesting quantum corrections to the explicit expression of the
constraint and the transformations it generates.

\subsection{Canonical quantization}

The canonical form (\ref{Friedmann3}) makes the first step of quantization
clear: we introduce a wave function $\psi(a)$ and replace $p_a$ with a
derivative operator $\hat{p}_a=-i\hbar\partial/\partial a$. Depending on how
matter is realized, by some field $\phi$ or a phenomenological perfect fluid,
$\rho$ might turn into an operator as well. In this way, we derive from the
Friedmann equation the Wheeler--DeWitt equation \cite{DeWitt}
\begin{equation} \label{WdW}
 \frac{2\pi G\hbar^2}{3V_0}\frac{1}{a}\frac{\partial^2\psi}{\partial a^2} 
+V_0a^3\hat{\rho}\psi=0\,.
\end{equation}

Unlike standard quantum-mechanical Hamiltonians, the Friedmann constraint
does not lead to a unique ordering of the non-commuting operators $\hat{a}$
and $\hat{p}_a$. Another problem arises, the factor-ordering problem of
quantum cosmology. (Presumably, one may want to choose a symmetric ordering,
but this choice would be neither obvious nor unique.) More generally, the
Wheeler-DeWitt equation or its extensions in different approaches usually
suffer from a large dose of quantization ambiguity.

One does not expect the Wheeler--DeWitt equation to be the final form of a
state equation in quantum cosmology, even when all its possible quantization
ambiguities are taken into account. In addition to quantum corrections that a
state obeying (\ref{WdW}) would imply compared to classical variables subject
to (\ref{Friedmann3}), there may be modifications because the classical
picture of continuum space-time, with some line element such as (\ref{Line}),
no longer applies when large energy density or curvature is reached. Some
popular ones of these modifications are described below and in
Sec.~\ref{s:Many}. Nevertheless, at reasonably small curvature, which in
cosmology is realized for most, if not all, of the space-time of the
accessible universe, the Wheeler--DeWitt equation should be a useful first
approximation. It can shed light on general conceptual issues of quantum
cosmology, show how the semiclassical limit of the state equation can be
compatible with the classical dynamics, and indicate how potential quantum
corrections could influence the physics of the cosmos. For instance, the
equation has been used in several detailed studies of cosmic decoherence
\cite{QuantClassCosmo,GravitonEntropy,PointerInflation} and for potential
effects on structure formation
\cite{NonSingBohmQC,CosmoWithoutInfl,LargeUniverse,WdWCMB}.

\subsection{Test structures}
\label{s:Test}

In order to probe candidates of quantum gravity, one can introduce their most
characteristic ingredients as modifications to the Wheeler--DeWitt equation
(\ref{WdW}) and explore their consequences. Examples include new types of
matter fields or condensates \cite{Tachyon}, (super)symmetries
\cite{SUSYQuantCos,Moniz} or extra dimensions \cite{Roy} motivated by string
theory, non-standard space-time structures as they may arise in
non-commutative \cite{Connes,NonCommST} or fractal geometry \cite{Fractional},
or discretization effects as studied in loop quantum gravity, causal dynamical
triangulations, or related approaches.

Staying in the canonical setting, one may, as in the rather widely-studied
example of loop quantum cosmology \cite{LivRev,Springer}, try to implement
discrete geometrical structures by formulating the canonical pair $(a,p_a)$
not as belonging to a continuum of spatial geometries but rather to a variable
$a$ that will, upon quantization, acquire a discrete spectrum. The derivative
operator $\hat{p}_a$ is then replaced by a finite shift operator that changes
$a$ (or some function of it, such as the volume) by a smallest discrete
increment. The differential equation (\ref{WdW}), accordingly, turns into a
difference equation. Since finite differences can always be Taylor expanded
when variations of the wave function are small, (\ref{WdW}) is obtained as a
continuum limit, but discreteness corrections are present at high curvature
where $p_a$ (or some function of $a$ and $p_a$) is large.

For details, one must use additional ingredients suggested more or less by
some full theory of quantum gravity, such as loop quantum gravity
\cite{Rov,ThomasRev,ALRev} which motivated loop quantum cosmology. One must
know which function of $a$ has an equidistant spectrum after quantization, so
that a finite shift operator can act on it. One must know how large or small
the smallest (Planckian?)  increment should be. And one must know how exactly
the finite shift operator has to replace $\hat{p}_a$ in (\ref{WdW}). All these
questions are far from being answered uniquely, so that a vast set of new
quantization and discretization ambiguities arises. But still, most of these
discreteness modifications are quite characteristic and can show what new
phenomena could be implied in addition to the quantum effects already present
in the Wheeler--DeWitt equation.

One problem that all test structures, including discretization effects, must
face is the ``problem of re-quantization.'' In most cases, we do not know to
what extent these ingredients are truly effective, in the sense that they
could be derived in some low-energy or semiclassical limit of a general theory
of quantum gravity.  Some of the test structures themselves, such as the
discreteness of loop quantum cosmology or extra dimensions in string theory,
may be effective or coarse-grained descriptions of a more fundamental quantum
theory. If they are introduced in the Friedmann equation before or while it is
being quantized to a modified Wheeler--DeWitt equation, one would make the
mistake of quantizing a theory that already contains quantum
corrections. (Indeed, the modifications leading to finite shifts in loop
quantum cosmology are often assumed to depend on $\hbar$ via the Planck
length.) That such a procedure might give wrong results can already be seen
for the well-understood systems of anharmonic oscillators. For a
quantum-mechanical particle of mass $m$ in a potential $V(x)$,
$\hat{x}$-expectation values in semiclassical states evolve according to
``classical'' motion in an effective potential
\begin{equation} \label{EffPot} 
V_{\rm eff}(x)=V(x)+ \hbar\sqrt{\frac{V''(x)}{m}}+O(\hbar^2)
\end{equation}
to first order in the $\hbar$-expansion and in an adiabatic approximation
\cite{EffAcQM,EffAc}. (See Sec.~\ref{s:ECD}.) If motion in this potential is
re-quantized, one obtains, by inserting $V_{\rm eff}$ for $V$ in
(\ref{EffPot}) and truncating to the same order in $\hbar$, motion in a
potential (\ref{EffPot}) with twice the square-root term. An iteration of the
procedure can produce arbitrarily large (but wrong) quantum corrections.  For
this reason, combined with the problem of quantization ambiguities, one cannot
make quantitative predictions in quantum cosmology until its equations have
been derived from some full theory of quantum gravity or the re-quantization
problem is solved in another way. But again, some effects are characteristic
enough to allow interesting qualitative investigations.

\subsection{Minisuperspace vs.\ single-patch}

If space is exactly homogeneous, a quantization of some region of coordinate
volume $V_0$ or geometrical volume $V_0a^3$ can be applied to any region with
non-zero volume. In fact, no prediction of physical effects or observables
made with such a model can depend on the value of the single parameter
$V_0$. Homogeneity implies that it does not matter where in space the region
is located, and by coordinate invariance the magnitude of $V_0$ cannot play a
role: The value could be changed in two ways, by enlarging or shrinking the
region, or by rescaling the coordinates used; since general covariance implies
that coordinate choices have no effect on predictions, the value of $V_0$ does
not affect the homogeneous dynamics and can appear only in auxiliary
constructs. Making sure that it cancels in the final expressions for
predictions is an important consistency condition for models of quantum
cosmology.

The Wheeler--DeWitt equation (\ref{WdW}) does not depend on $V_0$ or the
non-invariant scale factor $a$ separately, but only on the combination
$V_0a^3$. It is therefore rescaling invariant for a fixed homogeneous region
${\cal V}$ of volume $V=V_0a^3$. Different choices of homogeneous regions, one
with coordinate volume $V_0$ and one with $\lambda V_0$, lead to different
wave functions: $\psi(V)$ compared with $\psi(\lambda V)$. But since there is
no absolute scale to compare $V$ with, the $\lambda$-factor is not
problematic.  This conclusion might no longer be realized when one extends the
quantization by other quantum-gravity effects. In particular the discreteness
corrections of some models bring in a new scale, such as the Planck length. If
one implements discreteness by a difference equation of constant Planckian
step-size in the volume $V=V_0a^3$, for instance, terms such as
$\psi(V+n\ell_{\rm P}^3)$ would not simply scale under changes of $V_0$.

As in this example, quantum space-time effects may make it difficult to ensure
covariance under changing $V_0$. It may then be tempting to turn this volume
into a physical parameter,\footnote{Alternatively, it has been suggested that
  $V_0$ should be taken to infinity after quantization, interpreting it as
  some kind of infrared regulator. However, since the value of $V_0$ does not
  affect classical observables at all, it is not a regulator. Moreover, it is
  then difficult to reconcile models with compact and non-compact spaces
  within the same setting.}  by requiring it to be the actual size of some
distinguished region in the universe, for instance all that is accessible
within the Hubble distance. Unfortunately, there are several problems with
this view. First, the Hubble region is not an actual spatial region in the
universe. It is rather part of the past light cone because it is defined by
electromagnetic (or perhaps gravitational-wave) signals used for
observations. Secondly, the classical Friedmann equation does not require one
to make a distinguished choice for the region ${\cal V}$, and so it would seem
puzzling if some quantum models would force one to select a classical
region. Since these problems are usually caused by bringing in a new scale,
for instance by discreteness, the only distinguished region would be one of
just this size, for instance a region of the discreteness size --- a true
quantum effect --- or a smallest possible chunk of space that cannot be
further subdivided. No such limit exists in the classical continuum, and
therefore it is not surprising that no distinguished region appears in the
classical equation, while it shows up after bringing in quantum effects.

Traditionally, the Wheeler--DeWitt equation has been formulated as an equation
quantizing the behavior of all of space at once. The wave function $\psi(a)$
or $\psi(V)$, accordingly, has been identified as the ``wave function of the
universe.''  Within Wheeler--DeWitt quantum cosmology, this view is formally
consistent because the Wheeler--DeWitt equation is compatible with the
classical rescaling covariance. New quantum effects, however, lead to a more
complicated view and require one to change the picture. In discrete theories
it is more meaningful to view the quantized region as a smallest discrete
chunk of space, or a spatial atom. Not only problems with any scaling
dependence can be solved, but the picture also gets closer to familiar
physical systems in which one first quantizes a single atomic building block
and then asks how different ones interact. We take this new viewpoint in the
present review, and explore it further in the next section.

Before doing so, it is useful to note additional features of this change in
viewpoint. If one interprets the Wheeler--DeWitt equation as one quantizing a
whole universe, there is a big leap to take when exact homogeneity is relaxed
to even just a tiny bit of perturbative inhomogeneity. Many such attempts have
been made and are still being made, thanks to the important role of
cosmological perturbation theory for structure formation in the early
universe. Most of these constructions, however, run into some kind of
inconsistency related to the problem of covariance. In the isotropic context,
by taking a time derivative one can combine the Friedmann equation with the
continuity equation $\dot{\rho}+3(\dot{a}/a)(\rho+p)=0$ of matter with
pressure $p$ to obtain the second-order Raychaudhuri equation. The latter also
follows from the spatial part of Einstein's equation, which is then
automatically consistent with the physical requirement of energy
conservation. (In general, this is a consequence of the Bianchi identity of
the Einstein tensor.) The combination of evolution equations with energy
conservation is much more non-trivial when not only time but also space
derivatives must be taken into account for an inhomogeneous geometry. While
the classical field equations are still consistent thanks to the Bianchi
identity, this feature is not guaranteed when the equations are naively
modified by some suspected quantum effects. Instead, one must work hard to
show internal consistency and covariance of the quantum theory.  No internal
consistency condition need be imposed for the Wheeler--DeWitt equation of
homogeneous models to be meaningful, but any kind of inhomogeneity requires
delicate constructions to ensure that the resulting field equations are
consistent and covariant.

A varying, inhomogeneous field behaves in a much more complicated way under
spatial coordinate transformations than a constant function which would not
change at all. Homogeneous models are blind to spatial-covariance requirements
and only show invariance under reparameterizations of time. Time is delicate
too, as shown by the problem of time which is present even for homogeneous
models, but its 1-parameter transformations can be dealt with in a more direct
way than covariance under changes of all space-time coordinates. The
traditional view on quantum cosmology hides these problems, unlike the atomic
patch-picture which always makes it clear that at all levels there is some
structure and no exact homogeneity.

\section{Patching models}
\label{s:Patching}

The concept of a nearly homogeneous region of space splitting up into multiple
parts has appeared in several different forms. It not only assists in
simplifying the complicated non-linear dynamics of general relativity in a
more local setting in which interactions of the geometry on different regions
play only a secondary role, it also prepares the ground for an application of
notions and methods of condensed-matter physics in classical or quantum
cosmology.

Classical versions of this concept usually take the form of fractionalization
of an evolving spatial geometry into independent parts, and can be used in
different regimes. At low curvature, the separate-universe approach has been
useful in the study of structure formation in the early universe. At high
curvature, the Belinski--Khalatnikov--Lifshitz scenario has been influential
in understanding the approach to a spacelike singularity such as the one at
the big bang.

Quantum versions are more recent and are technically related to Bose--Einstein
condensation. Although there is no physical condensation of constituents
occupying the same state, near-homogeneity of a spatial geometry allows one to
think of a collection of patches being in almost the same state. The
multi-variable dynamics of inhomogeneity (or an interacting many-patch system)
is then modeled by a non-linear wave equation for a single patch, akin to the
Gross--Pitaevski equation.

\subsection{Fractionalization}

As a useful approximation in cosmology, the separate-universe approach
\cite{SeparateUniverse,SeparateUniverseII,SeparateUniverseIII} decomposes
space-time into small patches, depending on the wave length of modes
considered. To lowest order in the approximation, each patch can be treated as
an independent isotropic model with simple Friedmann dynamics. While this
picture resembles features of the view espoused here for quantum cosmology,
the scales are different: they are purely classical in the separate-universe
approach while they are quantum and microscopic in quantum cosmology.
Nevertheless, this approach is amenable to a quantum-gravity formulation, as
given for instance in \cite{SeparateUniverse}.

The Belinski--Khalatnikov--Lifshitz scenario \cite{BKL}, by contrast, works in
regimes of high curvature and strong inhomogeneity, but is still able to use
nearly-homogeneous dynamics at a very local level. It has been shown in
several models that time derivatives in Einstein's equation dominate spatial
derivatives when curvature is large
\cite{NumSing,Garfinkle,WS:AR,PastAttract}. The latter can then be ignored,
leaving an evolution equation according to which each point changes its
geometry independently of its neighbors. The scenario is asymptotic, so that
there is no general estimate on how close to a singularity it becomes a good
approximation. Nor is there an estimate on the size of regions that can, for a
given average curvature, be considered homogeneous. Also here, since the
scenario is classical, the sizes are in general unrelated to the Planck
scale. Nevertheless, the scenario supports the idea of patching of homogeneous
chunks of space even in regimes far from actual homogeneity.

The original Belinski--Khalatnikov--Lifshitz scenario is purely classical,
based on an analysis of Einstein's equation. But it can sometimes be
strengthened or made more general by including certain quantum effects. One
example is the introduction of supersymmetry and the powerful mathematical
tools accompanying it \cite{Billiards}. And also some modifications of
space-time dynamics suggested by spatial discreteness help to enhance time
derivatives over spatial ones \cite{NoSing}.

\subsection{Condensation}

If a nearly homogeneous collection of $n$ patches is uncorrelated and has all
constituents in the same state $\psi$, the total state is $\Psi=\psi^{\otimes
  n}$. Approximately, this quantum system with interaction potential $\hat{W}$
can then be described by an effective potential
$\langle\Psi|\hat{W}|\Psi\rangle$. For delta-function interactions
$W_{\alpha}(x_1,\ldots,x_n)=\frac{1}{2}\alpha\sum_{i\not=j} \delta(x_i-x_j)$
of pointlike particles, for instance, one obtains
\begin{equation}
 \langle\Psi|\hat{W}_{\alpha}|\Psi\rangle=\frac{1}{2}\alpha (n-1) \sum_{i=1}^n
 \int |\psi(x_i)|^4 {\rm d}x_i \,.
\end{equation}
The effective many-body potential therefore takes the form of the expectation
value of a sum of non-interacting but non-linear, wave-function dependent
``potentials'' $\frac{1}{2}\alpha (n-1) |\psi(x)|^2$. If this potential is
formally inserted in a 1-particle Schr\"odinger equation, one obtains the
non-linear wave equation
\begin{equation}
 i\hbar\frac{\partial\psi}{\partial t}=
 -\frac{\hbar^2}{2m}\frac{\partial^2\psi}{\partial x^2}+ \frac{1}{2}\alpha
 (n-1) |\psi(x)|^2 \psi(x)
\end{equation}
for each particle.  A rigorous derivation has been given
\cite{KineticEqs,NonLinSchroed}.

For patches of a spacelike geometry, the interaction potential, obtained from
Einstein's equation, is not pointlike. It is a polynomial in the metric and
its spatial derivatives, which latter can be approximated as finite
differences of neighboring patch geometries. An effective condensate potential
can still be derived, and it remains non-linear. However, owing to the
non-pointlike nature of interactions, it is non-local \cite{NonLinLQC}. It can
be expressed in terms of quantum fluctuations or higher moments of a
state. For instance, for quadratic interaction potentials of patch volumes
$V_i$, $\langle\Psi|\hat{W}|\Psi\rangle$ is a sum of terms $\int{\rm d}V_1{\rm
  d} V_2 |\psi(V_1)|^2|\psi(V_2)|^2 (V_1-V_2)^2=\int{\rm
  d}V|\psi(V)|^2\int{\rm d}v |\psi(V+v)|^2v^2$ and can for sharply peaked
states be approximated as the volume fluctuation $(\Delta V)^2$. The
corresponding Gross--Pitaevski-type equation is then non-local (and
non-linear) because it has a potential $(\Delta V)^2$ that depends on the
values of the wave fuction everywhere on configuration space.  Related
equations have also been obtained with more input from a full proposal for
quantum gravity \cite{GFTCosmo,GFTCosmo2,GFTCosmo3}. Such non-local equations
are more difficult to analyze, but there are still some simplifications
compared to any general many-patch equation as it may follow from a theory of
quantum gravity.

As a general paradigm, we can see quantum cosmology as an approximate
condensate of quantum gravity. A detailed realization of this picture can give
access to the homogeneous background dynamics as well as inhomogeneous
perturbations for structure formation. Several deep problems must, however, be
solved. We must know what states $\Psi$ and what form of operators
$\hat{W}$ are meaningful, and what consistency conditions they must
obey. As we will describe in the next section, different proposals can be
extracted from existing candidates for quantum gravity, but there is no firm
or even unique answer.

\section{Quantum gravity: Fundamental theories for many-patch systems}
\label{s:Many}

One of the difficulties associated with quantum gravity is the fact that the
candidates suggested for fundamental mathematical structures are rather far
removed from the familiar classical space-time picture or even from simple
constituent models such as patch systems. It is therefore hard to compute
physical phenomena or predictions from these ``first'' principles. All the
physical intuition, necessary to tell which approaches are promising and to
develop them further, is extracted after some extrapolations and heuristic
steps. Although the resulting models share characteristic features with the
fundamental formulations, a precise relationship is often lacking. In a few
cases, there are simple and solvable toy models which exhibit the expected
effects. But it remains unclear what happens under more general (and more
realistic) circumstances that do not provide complete mathematical
control. The situation is similar to complicated condensed-matter effects
which cannot yet be derived from an interacting many-body Hamiltonian. In
quantum gravity one uses a similar strategy, trying to find good effective
theories that may not be firmly derived but still capture interesting
phenomena. The crucial difference to condensed-matter physics is that
observations do not help much to pin-point the right effective equations.

In this section we briefly review the main features of the most popular
candidates for quantum gravity, followed in the next section with a discussion
of possible consequences at an effective level. The general description must
necessarily be incomplete, but hopefully gives an idea of the diversity of
different proposals.

\subsection{String theory}

String theory \cite{String} begins with the dynamics of a test string moving
and vibrating in classical space-time. After quantization, one obtains an
interacting tower of oscillator modes that may produce properties of known
elementary particles (and more). The modes include excitations suitable for a
description of quantized gravitational waves, and therefore the theory
promises to capture not just the quantum physics of matter but also of
space-time. In order to understand characteristic properties of quantum
space-time in this setting, it is important to control the rich and
complicated structure of non-perturbative phenomena. If this can be achieved,
full quantum space-time rather than just small excitations about it can be
analyzed.

Much is known about non-perturbative effects in string theory. However, in
order to derive them, one makes ample use of dualities among the different
regimes and descriptions of the theory, most prominently of the AdS/CFT
correspondence. At this stage, it becomes more difficult to derive a direct
space-time picture of fundamental effects, although models that resemble for
instance some types of topological defects \cite{Roy} or different kinds of
condensates of fundamental excitations
\cite{Tachyon,UniverseExpandString,InfoUniverse} are available.

\subsection{Non-commutative and fractal geometry}

Much of the shape geometry of a 2-dimensional surface can be extracted from
its vibration spectrum. By analogy, one may try to describe space-time
geometry by the spectrum of suitable differential operators that describe
waves in space-time. A convenient choice is the Dirac operator
$D=\slash\!\!\!\partial= i\hbar^{-1}\gamma^{\mu}\partial_{\mu}$ that
determines how fermions or spinors, on whose components the matrices
$\gamma^{\mu}$ act, propagate.  An analog system can be seen in a graphene
layer, on which phonons propagate by comparable laws.

Rather surprisingly, the abstract Dirac operator is related to gravity because
the ``spectral action,'' defined as the sum of all its eigenvalues up to some
(Planckian) maximum, turns out to produce the Einstein--Hilbert action in the
low-curvature limit \cite{DeWittDyn,Seeley}. One may therefore hope that
higher-order corrections in the curvature expansion, or perhaps the full
non-perturbative spectral action, can bring one closer to a quantum theory of
gravity.

An interesting feature is that the Dirac operators can be generalized to
non-classical manifolds, in particular to those on which functions do not
commute under multiplication \cite{Connes}. An indirect description of a
possible version of quantum, or non-commutative, space-time is then
obtained. However, an outstanding problem in non-commutative geometry is a
proper formulation for space-time, as opposed to 4-dimensional timeless
(Euclidean) spaces as commonly used. The problem of time therefore does not
play a role, but only because in this setting time remains poorly understood
at an even more basic level.

Fractal or fractional \cite{Fractional} spaces have been proposed as an
alternative deviation from the classical continuum. These effects change the
cosmological dynamics, in which context the proposal has mainly been explored
\cite{FractCosmo}.

\subsection{Wheeler--DeWitt quantum gravity}

Like string theory, canonical quantum gravity starts with a controlled
space-time picture. The central idea is to apply standard quantization methods
to quantities that describe the geometry and curvature of space-time. As
already seen in the Wheeler--DeWitt quantization of cosmological models, a
possible starting point is the metric $h_{ab}$ that determines the geometry of
space. Now, however, one allows for arbitrary position-dependent symmetric
matrices $h_{ab}$, rather than just isotropic ones that are determined by a
single time-dependent function $a(t)$. The role of the momenta $p^{ab}$ of
these variables is then played by a linear combination of matrix elements
$\dot{h}_{ab}$, geometrically identified as the extrinsic curvature of spatial
hypersurfaces in space-time \cite{ADM,ADMRe}. (The necessity of taking a time
derivative again suggests subtleties related to covariance since the choice of
coordinates should not matter for physics but would affect the form of
$\dot{h}_{ab}$. We come back to this issue in the next section.)

Having fields and their momenta, one can attempt a canonical quantization by
postulating wave functionals $\psi[h_{ab}]$ on the space of all metrics, on
which $p^{ab}$ would become some functional derivative \cite{DeWitt}. One
would have to make mathematical sense of these operations, using methods from
quantum field theory. With a well-defined Hilbert-space setting of
Wheeler--DeWitt quantum gravity one would be close to deriving an effective
picture of quantum space: An effective metric could be obtained from
expectation values of the metric operator in suitable semiclassical
states. Unfortunately, however, it has proven difficult to define, in analogy
with quantum mechanics, an inner product by some kind of integration over the
unwieldy and infinite-dimensional space of all metrics.

Quantum field theory routinely deals with infinite-dimensional configuration
spaces, but its standard techniques are not applicable in this context: if one
tries to construct well-defined operators, one is led to integrations of the
basic fields, for instance when one turns classical modes into ladder
operators. Such integrations or the definition of modes require a metric for
them to be well-defined, but the metric is what we are trying to quantize
here. There are two obvious ways out, none of which is promising: One could
use an auxiliary metric for the integrations, but its choice will then affect
properties of the physical metric operators. Or one could use the same
physical metric $h_{ab}$ to define integrations, but then the resulting
operators will obey a complicated non-linear algebra which makes it difficult
to analyze excited states. Wheeler--DeWitt quantum gravity has therefore
stayed at a formal level, but it has led to some results about possible
semiclassical behaviors.

\subsection{Loop quantum gravity}

Loop quantum gravity \cite{Rov,ThomasRev,ALRev} suggests a solution to the
Hilbert-space problem of Wheeler--DeWitt quantum gravity by using different
fields to describe the geometry. Instead of the metric one uses a densitized
triad $E^a_i$ (so that $E^a_iE^{bi}=h^{ab} \det(h_{cd})$, summed over the
repeated index $i$), a triple of orthonormal vector fields at each point in
space. Its momentum can be defined in the form of a connection $A_a^i$
\cite{AshVar,AshVarReell}, or a non-Abelian version of the familar
electromagnetic vector potential. Like the vector potential, $A_a^i$ can be
integrated over curves in space (the eponymous loops) without requiring an
auxiliary metric, and similarly the fields $E^a_i$ can be integrated over
surfaces in space to obtain an analog of electric flux. These integrated
objects can be quantized in a well-defined and manageable way \cite{FluxRep},
providing powerful methods to construct the quantum state space \cite{ALMMT}.

In quantum field theory, one follows the harmonic-oscillator model of
constructing state spaces by ladder operators. An operator $\hat{a}^{\dagger}$
that commutes with the energy operator $\hat{E}$ via
$[\hat{E},\hat{a}^{\dagger}]=\hbar\omega\hat{a}^{\dagger}$ raises the energy
level when it acts on states, by the familiar argument:
$\hat{E}(\hat{a}^{\dagger}|n\rangle)=
(\hat{a}^{\dagger}\hat{E}+[\hat{E},\hat{a}^{\dagger}])|n\rangle=
(E_n+\hbar\omega) \hat{a}^{\dagger}|n\rangle$. The observer-dependent notion
of energy is not fundamental in quantum gravity, and one rather refers to
geometrical quantities such as areas and volumes. The values of the triad
$E^a_i$ are closely related to areas. Ladder operators should therefore be
constructed from expressions $h$ that enjoy commutator relationships of the
form $[\hat{E}^a_i,\hat{h}]\propto \hat{h}$. Since $A_a^i$ is canonically
conjugate to $E^a_i$, a suitable expression for $h$ takes the form
$h_{\ell}(A)\sim \exp(i\ell A)$, or, being more careful with indices, equals
the holonomy
\begin{equation} \label{hol}
 h_{\ell}(A)={\cal P}\exp(i\smallint_{\ell}t^aA_a^j\tau_j{\rm
  d}\lambda)
\end{equation}
where $2i\tau_j$ are Pauli matrices, $\ell$ is now a curve in
space with tangent vector $t^a$, and ${\cal P}$ denotes a suitable
ordering of non-commuting SU(2)-elements along the curve.

Using these ladder operators, one is directly led to a discrete picture of
space, in which the loops of the integrated $A_a^i$ create geometrical
excitations only along 1-dimensional curves or within a
graph. Correspondingly, one can show that the integrated $E^a_i$ are quantized
to operators with discrete spectra \cite{AreaVol,Area,Vol2}. Since $E^a_i$ is
the substitute for $h_{ab}$ as a measure of spatial geometry, space acquires a
discrete structure.
It is possible to turn the full Hamiltonian constraint into a well-defined
operator \cite{AnoFree,QSDI}, an important result which has remained elusive
in the more formal Wheeler--DeWitt approach.

The theory is non-perturbative from the outset, and yet, unlike in string
theory, there is an appealing picture of quantum space. Unfortunately,
however, the transition to quantum space-{\em time} is much less tidy. Not
only is the dynamics of these discrete spaces forbiddingly complicated,
resembling an interacting spin system on an irregular and changing
lattice. Also the problem of time plays a key but still poorly understood
role, for one has to ensure that the choice of time coordinate in the
definition of momenta does not affect the physics. The underlying classical
symmetry can easily be broken by quantum anomalies, which would render
meaningless any given proposal for the dynamics which turns out to be
anomalous.

\subsection{Discrete covariant approaches}

There are different versions of discrete theories which, unlike loop quantum
gravity, introduce their structures directly for space-time. The most advanced
among them is the theory of causal dynamical triangulations \cite{DiscQG}, in
which one introduces discretized space-time as a fundamental theory and looks
for a second-order phase transition whose long-range correlations would give
rise to a corresponding continuum theory.
An influential result in this framework has been the demonstration that its
quantum space-time structure differs from the classical one on small
distances, expressed by a reduction of the dimension to values smaller than
four \cite{LowDimDynTriag}. 

As a different approach, spin-foam models
\cite{SurfaceSum,AlRo,PerezLivRev,EngleSpinFoams} have been suggested as an
``evolving-graph'' (or path-integral) version of loop quantum gravity. The
1-dimensional structures that describe quantum space should then evolve by
moving about and branching out when new curves are created. The discrete
space-time picture is therefore 2-dimensional or foam-like, with sheets swept
out by moving loops connected along lines of branching events. Although there
is no obvious place where a time coordinate is chosen, in contrast to
canonical theories, the dynamics of such models and the question whether
space-time symmetries are properly realized remain poorly understood. The
intuitive space-time picture provided by these models is hard to control, with
most calculations so far restricted to toy systems with a small number of
links.

As a possible way to manage spin foams at a more general level, it has been
suggested to use field theories on a group manifold whose Feynman diagrams in
a perturbative treatment are identical to spin foams. These group-field
theories \cite{GFT} have evolved into an independent approach, viewed as a
possible second quantization of quantum gravity.

\subsection{Asymptotic safety}

One could forego complicated fundamental structures if one could show that the
non-renormalizable perturbative quantization of gravity becomes meaningful
when expanded around a non-trivial vacuum. 
The classically suggested vacuum may turn out to be unsuitable for an
expansion of the theory, for instance when the classical coupling constants
are unstable under the renormalization flow. This flow may nevertheless have
fixed points for which some or all coupling constants are large, so that an
expansion around these values would be better behaved.

In order to test this proposal of asymptotic
safety \cite{AsSafe}, the renormalization-group flow of classical general
relativity has been studied in some detail, providing promising indications
that a non-Gaussian fixed point may indeed exist. 
This problem is complicated because the flow away from small couplings
progressively magnifies the importance of higher-curvature terms in an
effective action for gravity. Coefficients of the latter play the role of new
coupling constants, leading to a large parameter space to be probed.
Current problems of this approach are related to the fact that one must
considerably truncate the large theory space of all curvature-dependent
actions in order to produce manageable models, as well as the fact that
different formulations and choices of fields do not always seem to produce the
same results.

\section{Effective theories}
\label{s:Eff}

Actual derivations of physical effects remain challenging in all approaches to
quantum gravity, at conceptual and computational levels. But it is usually
possible to parameterize characteristic effects, such as the topological
defects of string theory or the discreteness of loop quantum gravity, and
evaluate them in simple models in order to put those theories to the
test. Such results can hardly be justified as predictions of the underlying
theories, but they may still serve as a means to falsify some radical
proposals and to reduce the number of choices and ambiguities.

\subsection{Branes and other worlds}

There are several different classes of effective models of string effects,
which are usually obtained by including topological structures or additional
fields in extended versions of Einstein's equation for the space-time metric
(see for instance \cite{StringCosmoGio,Roy,StringCosmo}). These modifications
are suggested either by excitations of a string or from non-perturbative
effects. A full derivation as an effective theory descending from the
fundamental one is still lacking. Such models are rarely formulated as systems
of quantum cosmology but rather as models of the classical type, with a
continuous space-time structure subject to laws with quantum
corrections. Therefore, the re-quantization problem is avoided, and the
problems of time, covariance and states do not play a role.

\subsection{Non-commutative geometry}

A simple model of non-commutative geometry is obtained if one replaces the
commutative algebra of functions on space-time by non-commuting
$2\times2$-matrices. Schematically, one can then write the new Dirac operator
as $\hat{D}=\left(\begin{array}{cc}\slash\!\!\!\partial &
    S\\S&\slash\!\!\!\partial\end{array}\right)$ with another operator $S$.
(The single graphene layer of an analog model for the Dirac operator then
becomes a bi-layer of interacting sheets of graphene.) Since $S$ may be
position-dependent, it provides new fields of potential interest for particle
or cosmological phenomenology \cite{NonCommStandard}. Another effective
approach is to extend the curvature expansion of the spectral action to higher
orders, so that possible physical consequences can be studied at the level of
higher-curvature corrections \cite{NonCommCosmo}.

\subsection{Effective canonical dynamics}
\label{s:ECD}

The most systematic approach to effective theories exists in canonical quantum
gravity. The formalism starts with the familiar Ehrenfest theorem of quantum
mechanics, but includes several new features to take into account the
requirements of space-time theories, such as general covariance or allowing
for the freedom to change the time coordinate. Canonical methods are best
suited for these extensions of familiar effective descriptions.

By Ehrenfest's theorem, expectation values of position $q$ and momentum $p$ of
a single particle of mass $m$ in a potential $V(q)$ change in time by
\begin{equation}
 \frac{{\rm d}\langle\hat{q}\rangle}{{\rm d}t} =
 \frac{\langle[\hat{q},\hat{H}]\rangle}{i\hbar}= 
   \frac{\langle\hat{p}\rangle}{m} \quad,\quad
 \frac{{\rm d}\langle\hat{p}\rangle}{{\rm d}t} = -\langle
   V'(\hat{q})\rangle\,. \label{qp}
\end{equation}
These equations closely resemble the classical ones, except that they couple
to each other by the expectation value $-\langle V'(\hat{q})\rangle$ of the
force, rather than the force function $-V'(\langle\hat{q}\rangle)$ evaluated
in the $q$-expectation value. The difference of these two expressions gives
rise to quantum corrections.

\subsubsection{Moment dynamics.}

In a more systematic treatment, one expands the quantum force $-\langle
V'(\hat{q})\rangle=-V'(\langle\hat{q}\rangle)-\frac{1}{2}
V'''(\langle\hat{q}\rangle) \Delta(q^2)+\cdots $ in moments $\Delta(q^a) =
\langle(\hat{q}-\langle\hat{q}\rangle)^a\rangle$ of the state
considered. These moments generally change in time as the state spreads, so
that the system (\ref{qp}) must be extended by equations of motion for
$\Delta(q^a)$. Given the Hamiltonian operator
$\hat{H}=\hat{p}^2/2m+V(\hat{q})$, one can compute these equations in the same
way in which one proves Ehrenfest's theorem: Starting with the first
non-trivial choice $a=2$ (for which $\Delta(q^2)=(\Delta q)^2$ is the
$q$-fluctuation squared), one writes
\begin{equation}
 \frac{{\rm d}\Delta(q^2)}{{\rm d}t} =  \frac{{\rm
     d}(\langle\hat{q}^2\rangle- \langle\hat{q}\rangle^2)}{{\rm d}t}
=
 \frac{\langle[\hat{q}^2,\hat{H}]\rangle}{i\hbar}- 2
 \langle\hat{q}\rangle\frac{{\rm d}\langle\hat{q}\rangle}{{\rm d}t}
= 2\frac{\Delta(qp)}{m}\,, \label{dDq}
\end{equation}
introducing the covariance $\Delta(qp)$ as an example of the more-general
moments
\begin{equation} \label{D}
 \Delta(q^ap^b) =
 \langle(\hat{q}-\langle\hat{q}\rangle)^a(\hat{p}-
\langle\hat{p}\rangle)^b\rangle_{\rm
   Weyl} 
\end{equation}
with totally symmetric (or Weyl) ordering.

Proceeding in this way, the whole set of infinitely many moments
$\Delta(q^ap^b)$ is in general coupled to the expectation values. One can
approximate this version of quantum evolution by truncating the set of moments
so that only those with $a+b\leq n$ for some fixed $n$ are considered. A
finite set of coupled equations is then obtained. The truncation is suitable
for a semiclassical approximation because semiclassical states obey
$\Delta(q^ap^b)\sim O(\hbar^{(a+b)/2})$, but some other regimes are accessible
as well.
For instance, near-ground states of perturbed harmonic oscillators (or
perturbed free theories) can be accessible because the harmonic-oscillator
ground state is Gaussian (but not considered semiclassical). Another example
is given by properties of generalized uncertainty principles derived for phase
spaces with non-trivial topology \cite{MomentGUP}.

The coupled dynamics of all moments can be solved partially if one assumes
that moments change slowly. An adiabatic approximation can then be combined
with the semiclassical one \cite{EffAc}, with results equal to those derived
from the low-energy effective action known from particle physics
\cite{EffAcQM}. At lowest order, an expansion around the harmonic oscillator
with frequency $\omega$ produces the equations ${\rm
  d}\langle\hat{q}\rangle/{\rm d}t= \langle\hat{p}\rangle/m$ and
\begin{equation}
\frac{{\rm d}\langle\hat{p}\rangle}{{\rm
    d}t}=-m\omega^2\langle\hat{q}\rangle
-U'(\langle\hat{q}\rangle)-\frac{1}{2}
U'''(\langle\hat{q}\rangle)\Delta(q^2)+ \cdots \label{pdot}
\end{equation}
for expectation values, we well as (\ref{dDq}) together with
\begin{eqnarray}
\frac{{\rm d}\Delta(qp)}{{\rm d}t}&=&
\frac{1}{m} \Delta(p^2)-
m\omega^2\left(1+\frac{U''(\langle\hat{q}\rangle)}{m\omega^2}\right)
\Delta(q^2) +\cdots \label{dDqp}\\ 
\frac{{\rm d} \Delta(p^2)}{{\rm d}t}&=& -2m\omega^2
\left(1+\frac{U''(\langle\hat{q}\rangle)}{m\omega^2}\right)
\Delta(qp)+\cdots \label{dDp}
\end{eqnarray}
with anharmonicity $U(q)=V(q)-\frac{1}{2}m\omega^2q^2$. To leading adiabatic
order (indicated by a subscript zero), (\ref{dDq}), (\ref{dDqp}) and
(\ref{dDp}) are zero and one finds $\Delta_0(qp)=0$ and $\Delta_0(p^2)=
m^2\omega^2 (1+U''(\langle\hat{q}\rangle)/m\omega^2) \Delta_0(q^2)$. The
first-order equations (such as ${\rm d}\Delta_0(q^2)/{\rm d}t=
2\Delta_1(qp)/m$) then produce a consistency condition
solved by
$\Delta_0(q^2)=C_2/\sqrt{1+U''(\langle\hat{q}\rangle)/m\omega^2}$ with a
constant $C_2$. If the harmonic ground state is to be obtained for $U(q)=0$,
$C_2=\frac{1}{2}\hbar/m\omega$. Equation (\ref{pdot}), formulated for
$V(q)$, then shows that motion happens in the effective potential
(\ref{EffPot}).  To higher orders in the adiabatic approximation,
higher-derivative terms are added to the classical equations
\cite{HigherTime}.

\subsubsection{Deparameterization.}
\label{s:Deparam}

An application of these methods to quantum cosmology must take into account
the fact that the Friedmann equation is a constraint rather than an evolution
generator. A popular but problematic method often used in this context, called
deparameterization, consists in coupling a special choice of matter field to
space-time, so that its homogeneous value can formally play the role of
time. Its momentum is then the analog of the usual Hamiltonian, and standard
methods can be used to analyze quantum or effective evolution. Although the
method of deparameterization and the popular models obtained from it are
problematic, as we will see in more detail, we describe an example in order to
demonstrate some current technical developments in quantum cosmology but also
their pitfalls. For different versions, see
\cite{APS,BKDustLTB,DeparamQG,HusainDust}.

The most common example of this kind is a free, massless scalar field $\phi$
with momentum $p_{\phi}$, whose energy density $\rho=p_{\phi}^2/2a^6$
contributes to the Friedmann equation (\ref{Friedmann3}).  Solving $C=0$ for
the momentum $p_{\phi}$, we have
\begin{equation} \label{apa}
 p_{\phi}(a,p_a)= \sqrt{\frac{4\pi G}{3c^2}}\: |a p_a|\,.
\end{equation}
(We set $V_0=1$ from now on.) With this momentum interpreted as a Hamiltonian
for changes with respect to $\phi$ (assuming positive $ap_a$), we obtain
Hamiltonian equations of motion ${\rm d}a/{\rm d}\phi= \partial
p_{\phi}/\partial p_a= \sqrt{4\pi G/3c^2}\: a$ and ${\rm d}p_a/{\rm d}\phi=
-\partial p_{\phi}/\partial a= -\sqrt{4\pi G/3c^2}\: p_a$.  Both variables
therefore change exponentially with $\phi$.

Transforming to proper time by multiplying with ${\rm d}\phi/{\rm
  d}\tau= \partial C/\partial p_{\phi} = p_{\phi}/a^3$ with constant
$p_{\phi}$ (because ${\rm d}p_{\phi}/{\rm d}\tau= -\partial
C/\partial\phi=0$), one can see that the usual equations of motion follow:
$a^{-1}{\rm d}a/{\rm d}\tau=a^{-1}({\rm d}a/{\rm d}\phi)({\rm d}\phi/{\rm
  d}\tau)=\sqrt{4\pi G/3c^2}\: p_{\phi}/a^3$ is equivalent to the Friedmann
equation, and from ${\rm d}p_a/{\rm d}\phi$ one finds the Raychaudhuri
equation. Deparameterization therefore allows one to reformulate constrained
dynamics as a standard evolution picture, at the expense that one must provide
a specific form of matter field.

\subsubsection{Solvable model.}

The deparameterized model has a simple $\phi$-Hamiltonian (\ref{apa}) which
shares with the harmonic oscillator the fact that it is quadratic. The
Hamiltonian is not of the standard form with a kinetic term and a potential,
but one may still use the commutator version of Ehrenfest's equations
(\ref{qp}). As with the harmonic oscillator, one then finds that expectation
values of $a$ and $p_a$ obey exactly the classical equations after
quantization. There are no quantum corrections or effective potentials. The
model is very special, owing to the symmetry, the specific matter content, and
other properties such as the absence of spatial curvature or a cosmological
constant. Nevertheless, it is useful for studying the potential of quantum
space-time effects that might be implied by quantum gravity in addition to the
usual quantum back-reaction of moments.

Loop quantum cosmology, for instance, suggests that the quadratic curvature
dependence $p_a^2$ of the Friedmann equation is replaced by a periodic
function, as motivated by finite shifts in quantum geometry or matrix elements
of holonomies (\ref{hol}). After such a replacement of $p_a$ by
$\delta^{-1}\sin(\delta p_a)$ with some constant or perhaps $a$-dependent
$\delta$, the $\phi$-Hamiltonian
\begin{equation} \label{pmod}
 p_{\phi}(a,p_a)= \sqrt{\frac{4\pi G}{3c^2}}\delta^{-1} |a \sin(\delta p_a)|
\end{equation}
is no longer quadratic. Nevertheless, there is a hidden form of solvability
\cite{BouncePert} which one can realize after changing to variables $a$ and
$J:=\delta^{-1}a\exp(i\delta p_a)$. (In this transformation, $\delta$ is not
required to be constant but may depend on $a$ by a power law
\cite{BounceCohStates}. This generalization is of interest in the discussion
of scaling behaviors and the magnitude of quantum corrections \cite{APSII}.)

After quantization, these new variables obey the linear algebra
$[\hat{a},\hat{J}]= -\hbar \hat{J}$, $[\hat{a},\hat{J}^{\dagger}]= \hbar
\hat{J}^{\dagger}$ and $[\hat{J},\hat{J}^{\dagger}] = 2\hbar
\hat{a}$. Moreover, the $\phi$-Hamiltonian is linear in $\hat{J}$, being
proportional to $i(\hat{J}-\hat{J}^{\dagger})$. With these features,
Ehrenfest's equations for expectation values of $\hat{a}$ and $\hat{J}$ still
close, and no quantum corrections other than the modification appear. One can
easily solve these equations and find that the modified $a(\phi)$ does not
reach zero (it behaves cosh-like), avoiding the big-bang singularity: For
non-zero $a$, the energy density never diverges. (This solution is an example
for bounce cosmology. For other versions, see
\cite{BounceReview,CosmoWithoutInfl,PertBB}.)

These formulations are of mathematical interest and they show some quantum
effects, but they do not suffice to provide a reliable picture of the Planck
regime or the big bang. First, they suffer from the re-quantization problem:
One implements a modification of the Friedmann equation in the form
(\ref{pmod}) as motivated by properties of loop quantum gravity and its
geometry. The parameter $\delta$, referring to the step-size of discrete
geometry, then likely depends on $\hbar$, and is indeed often assumed to be
related to the Planck length. But then one quantizes this system which already
contains some quantum corrections, and the cautionary remark of
Sec.~\ref{s:Test} applies.\footnote{Loop quantum gravity has a well-defined
  regularization scheme which allows one to express the dynamics in terms of
  holonomies. The modification of loop quantum cosmology, as in (\ref{pmod}),
  however, is not a proper regularization because the limit $\delta\to0$
  cannot be taken after quantization. One can think of this crucial difference
  between the two frameworks as a consequence of some kind of averaging
  required to implement homogeneity. Since no complete derivation of
  cosmological models from the full theory is known one cannot pinpoint
  exactly where the first quantization step happens which leads to an
  $\hbar$-dependent $\delta$.}

A modification such as (\ref{pmod}) may still provide reliable qualitative
information about quantum corrections even if the re-quantization problem is
unsolved. However, the lack of quantitative control means that it is
impossible to tell whether the modification in (\ref{pmod}) is the dominant
quantum correction under all circumstances. The main other source of
corrections is, in general, quantum back-reaction to which we turn now.

There are no further quantum back-reaction terms that would couple moments to
expectation values in this particular model, but it is just as special as the
harmonic oscillator. With any additions to make the model more realistic,
quantum back-reaction results and corrections added to the modifications
appear, analogously to (\ref{pdot}). Owing to their rather involved
derivations, however, they are only incompletely known.  The second problem
with such models can then be seen: Based on general principles of effective
theory \cite{BurgessLivRev,EffectiveGR}, one does indeed expect quantum
corrections of the form of higher time derivatives, which in gravitational
theories are usually expressed as higher-curvature effective actions. These
corrections, on dimensional grounds, are expected to be of a tiny size given
by the average density divided by the Planck density. Corrections implied by
an expanded (\ref{pmod}) for small $\delta p_a$ can be estimated to be of
about the same order of magnitude if $c\delta a\sim\ell_{\rm P}$, so that one
cannot consistently use the whole series of the expanded $\sin(\delta p_a)$
while ignoring quantum back-reaction terms. For these reasons, the
high-density regime of modified equations in loop quantum cosmology remains
unclear.\footnote{Starting with \cite{APSII}, there have been several
  activities in minisuperspace-based loop quantum cosmology, aiming for a
  detailed understanding of wave functions subject to a dynamics quantizing
  (\ref{pmod}). Even formulating such a dynamical law is non-trivial because
  of certain conditions required for a so-called physical Hilbert space to be
  used. Moreover, some generalizations of (\ref{pmod}), especially those with
  a cosmological constant, do not automatically lead to self-adjoint quantum
  Hamiltonians \cite{NonSelfAd}. In spite of much progress in these
  investigations, including also the numerical front \cite{NumLQC,Chimera},
  the high-density behavior is uncertain. Most details, analytic or numerical,
  are available in solvable models quantizing (\ref{pmod}), and in fact most
  intuition in this field is based on the dynamics of kinetic-dominated models
  with vanishing or small quantum back-reaction. Numerical investigations
  sometimes attempt to go beyond kinetic domination, but they require specific
  choices of initial states and therefore lead straight to the problem of
  states to be discussed in more detail in
  Section~\ref{s:States}. Nevertheless, these different methods may eventually
  help to shed light on the generic behavior in the Planck regime of loop
  quantum cosmology.}  This is an example for the incompleteness problem of
quantum cosmology: Without knowing all terms of a general effective
description of quantum gravity, some of which are often difficult to derive,
one cannot make definitive predictions. Although such a statement may seem
close to a tautology, it is worth remembering in this field.

\subsection{The role of symmetries and space-time structure}

Space-time enjoys an infinitely large number of important symmetries. One of
the key problems of quantum gravity is to find consistent quantum realizations
of these transformations, much larger than the Euclidean group of rotations
and translations or even the Poincar\'e group that also contains Lorentz
boosts and time translations.

Poincar\'e transformations play an important role in particle-physics models
of quantum-field theory. They are generated by elements $P_{\mu}$ (space
translations for $\mu=1,2,3$ and time translations for $\mu=0$) and
$M_{\mu\nu}$ (rotations and boosts) with algebra relations
\begin{eqnarray}
  [P_{\mu},P_{\nu}] &=& 0\quad,\quad
  {} [M_{\mu\nu},P_{\rho}] = \eta_{\mu\rho}P_{\nu}-
  \eta_{\nu\rho}P_{\mu} \label{Poincare1}\\ 
  {} [M_{\mu\nu},M_{\rho\sigma}] &=& \eta_{\mu\rho}M_{\nu\sigma}-
  \eta_{\mu\sigma}M_{\nu\rho}- \eta_{\nu\rho}M_{\mu\sigma}+
  \eta_{\nu\sigma}M_{\mu\rho}\,.  \label{Poincare2}
\end{eqnarray}
These relations depend on the metric $\eta_{\mu\nu}$ of Minkowski
space-time. Since these matrix elements are just constants in Cartesian
coordinates, (\ref{Poincare1}) and (\ref{Poincare2}) define a Lie algebra.

\subsubsection{Deformations.}

The covariance symmetries of general space-times are an extension from
Poincar\'e transformations to arbitrary non-linear coordinate changes. Even
though space-time is one 4-dimensional entity, it is useful to split the
transformations in spatial and temporal changes, just as the Poincar\'e
algebra contains space and time translations as well as spatial rotations and
space-time boosts. One can then represent them geometrically as deformations
of spatial hypersurfaces within space-time, some tangential to hypersurfaces
and the rest normal. 

For every vector field $\vec{w}$ on a spatial hypersurface there is a spatial
deformation $S(\vec{w})$, and for every function $N$ on a spatial hypersurface
a normal (or timelike) deformation $T(N)$ which at each point $x$ shifts the
hypersurface by a distance $N(x)\vec{n}(x)$ with the unit normal vector
$\vec{n}(x)$ at $x$. By composing different such deformations, one obtains
algebraic relations that generalize the Poincar\'e algebra:
\begin{eqnarray}
 [S(\vec{w}_1),S(\vec{w}_2)]&=& S((\vec{w}_1\cdot\vec{\nabla})\vec{w}_2-
 (\vec{w}_2\cdot\vec{\nabla})\vec{w}_1) \label{DD}\\
{} [T(N),S(\vec{w})] &=& -T(\vec{w}\cdot\vec{\nabla}N) \label{HypDef}\\
{} [T(N_1),T(N_2)] &=& S(N_1\vec{\nabla}N_2-N_2\vec{\nabla}N_1)\,, \label{HH}
\end{eqnarray}
called the hypersurface-deformation algebra \cite{DiracHamGR}.  One can indeed
check that linear $N$ and $\vec{w}$ reproduce the Poincar\'e algebra.

The large number of hypersurface-deformation symmetries is a powerful input in
fundamental formulations of space-time theories. Any theory which is
formulated for the space-time metric, is invariant under these
transformations, and has the usual classical structure of second-order partial
differential equations must equal general relativity
\cite{Regained,LagrangianRegained}. Quantum effects could be realized only in
higher-order derivatives, by extra ingredients such as new fields, or by a
modification of the symmetries themselves. Studying hypersurface deformations
and their fate in any quantum theory of gravity therefore provides access not
just to the elementary space-time structure but also to its dynamics.

In contrast to the Poincar\'e algebra, the hypersurface-deformation algebra is
infinite-dimensional. Finding well-defined representations that could transfer
these important symmetries to a quantum theory of gravity is therefore
difficult. Even more importantly, the hypersurface-deformation ``algebra'' is
not an algebra in the strict sense. The gradient in the relation (\ref{HH})
requires one to use a metric on spatial hypersurfaces. Unlike the Minkowski
metric that appears in the Poincar\'e algebra, the spatial metric is not
constant under the general circumstances in which the hypersurface-deformation
symmetries are relevant. The spatial metric is therefore an external field,
independent of the symmetry generators and their coefficients $N$ and
$\vec{w}$, which must be included for a well-defined and closed formulation of
the symmetries. It turns out that Lie algebroids provide the right
generalizations of Lie algebras to take into account this extra ingredient
\cite{ConsAlgebroid}. However, while Lie algebroids have been studied
intensively in the mathematical literature of the recent decades, they are
harder to classify and to represent. Standard representation theory therefore
does not provide a simple road to quantum gravity.

\subsubsection{Representations.}

In physics, the first approach is often a brute-force construction rather than
elegant representation theory. Applied to the hypersurface-deformation
algebroid, this has been the attempt of canonical quantum gravity for several
decades, without much success. One of the key problems has been noticed
already in the late 1970s \cite{NonHerm,Komar}: The presence of the metric in
the algebroid relations leads to apparent contradictions between different
features that are usually taken for granted in quantum physics.

We are looking for a representation of $S(\vec{w})$ and $T(N)$ and their
relations by operators $\hat{S}(\vec{w})$ and $\hat{T}(N)$ acting on some
states $\psi$. States that obey the required symmetries and are invariant
under hypersurface deformations then solve the equations
$\hat{S}(\vec{w})\psi=0$ and $\hat{T}(N)\psi=0$ for all $\vec{w}$ and $N$. We
also expect that these symmetry generators are anti-Hermitian. (Their
exponentials should be unitary.) The left-hand side of (\ref{HH}) is then
anti-Hermitian, and so must be the right-hand side. The spatial generator
$\hat{S}(\vec{w})$, by our assumptions, is anti-Hermitian, but in quantum
gravity there is another operator for the metric $h_{ab}$ in the gradients. We
must order the right-hand side to something of the form
$\frac{1}{2}(\hat{S}\hat{h}+\hat{h}\hat{S})$. This ordering does not
annihilate an invariant state $\psi$ because the metric factor in
$\hat{S}\hat{h}\psi$ is not invariant under spatial deformations, while the
commutator on the left-hand side annihilates invariant states. The assumption
of anti-Hermition generators is therefore incompatible with a representation
of the algebroid and the existence of invariant states. (In path-integral
quantizations or the discrete version of spin-foam models, one does not
directly represent the generators as operators. Nonetheless, the
representation problem resurfaces as the anomaly problem of the path-integral
measure, which remains unresolved \cite{Anomaly,Secondary}.) 

One may relax the assumptions. There is no observable associated with the
generators. (For instance, on general space-times there is no invariant notion
of energy, as the usual observable related to time translations.) The
requirement of anti-Hermitian generators could therefore be weakened. However,
under these generalized circumstances some standard quantum notions no longer
apply, and one must carefully re-assess constructions of quantum field
theory. And even in this relaxed setting, it remains difficult to find
representations. An encouraging development in the last few years,
using the loop methods of \cite{AnoFree,QSDI},
has been the construction of consistent realizations in models with only two
spatial dimensions
\cite{ThreeDeform,TwoPlusOneDef,TwoPlusOneDef2,AnoFreeWeak}. An extension to
4-dimensional space-time is more complicated but may be feasible with these
new methods.

\subsubsection{Signature change.}

Effective methods have shown several unexpected implications of quantum
effects. In this setting, one does not directly compute commutators of
operators that quantize the generators in (\ref{HypDef}), but instead includes
quantum terms in the classical expressions of the generators and obtains their
commutator as a Poisson bracket \cite{ConstraintAlgebra}. Semiclassical
features of quantum space-time can then be uncovered.
Before presenting some of these consequences, it is useful to make two
remarks. First, owing to the complexity of the covariance or anomaly problem,
no complete picture of space-time structures in canonical quantum gravity is
available yet. The following discussion is based on a set of models which
appear to give a uniform and rather generic view on possible quantum
effects. Nevertheless, not all quantum corrections have been included in these
calculations. Secondly, since controlling inhomogeneity is an important
pre-requisite for a reliable analysis of potentially observable effects of
quantum gravity, several attempts have been made which circumvent an analysis
of the complicated anomaly problem. In loop quantum cosmology, for instance,
there are several such versions \cite{Hybrid,Extension} which {\em assume}
that the classical space-time structure (but not the classical dynamics) is
still realized in cosmological models of quantum gravity. These constructions
therefore amount to quantum-field theory on a (modified) background but do not
analyze full quantum space-time effects. 
The results described in the rest of this subsection are incomplete but
nonetheless indicate that classical space-time structures may not give the
full picture of the Planck regime in quantum gravity.

In particular, the characteristic features of loop quantum gravity lead to
quantum corrections not just in the dynamics of gravity but also in the
structure of space-time: The relation (\ref{HH}) is not realized in the
classical form but rather as
\begin{equation}
  [T(N_1),T(N_2)] = S(\beta
  (N_1\vec{\nabla}N_2-N_2\vec{\nabla}N_1))\,, \label{HHbeta} 
\end{equation}
with a function $\beta$ that depends on the metric or curvature
\cite{ConstraintAlgebra,JR,ScalarHol,LTBII,ModCollapse}. The functional form
of $\beta$ can be determined from the form of quantum corrections: Like
effective potentials or Hamiltonians, the effective generators are defined via
expectation values of the corresponding operators. The same kind of
corrections in the algebra can be seen by the mentioned operator calculations
\cite{ThreeDeform,TwoPlusOneDef,TwoPlusOneDef2,AnoFreeWeak} in models with
only two spatial dimensions.
(One could expect more general versions of the algebra in which also
(\ref{DD}) and (\ref{HypDef}) are modified, or even modified terms involving
the Gauss constraint in models based on triad variables. No concrete
examples have been found, and although they may be possible, they do not seem
to be likely because the diffeomorphism and Gauss constraint do not receive
strong corrections in loop quantum gravity.)

For $\beta\not=1$, the space-time structure is not classical: The symmetries
do not correspond to coordinate changes, and for linear $N$ and $\vec{w}$,
which classically produces the Poincar\'e algebra, modified transformations
are obtained \cite{DeformedRel,GeneralizedPoincare}. Nevertheless, the theory
is well-defined and physical observables can be computed from the generators
and the dynamics they produce. (In some cases, a canonical transformation
allows one to absorb $\beta$ in a re-defined spatial metric \cite{Absorb}, but
the space-time geometry remains non-classical:
An effective space-time obtained after the transformation would not be defined
for a direct quantization of the metric.)

The most interesting aspect of these non-classical space-time models is the
fact that $\beta$, with holonomy corrections from loop quantum gravity, turns
negative near the Planck density \cite{JR,ScalarHol}. One can show that this
sign change, in the linear limit, amounts to replacing boosts by rotations, so
that transformations of 4-dimensional Euclidean space are obtained instead of
those for space-time \cite{Action,PhysicsToday}. In other words, at high
density the time dimension turns into a spatial one, and equations of motion
generated by a system subject to (\ref{HHbeta}) are not deterministic.
Indeed, field equations computed in this regime take the form of elliptic
differential equations, not hyperbolic ones \cite{ScalarHol}. The usual
initial-value problem is to be replaced by a boundary-value problem in four
dimensions. (A possible analog model is a phase transition of nano-wires
\cite{SigChange}: There is no electric conductivity in the unordered phase,
but electricity --- just like time in quantum cosmology --- starts flowing
after the transition to an ordered phase.)  These consequences are an example
of the drastic changes to our intuitive notion of classical space-time that
can be implied by quantum effects. Any intuitive input necessary to formulate
quantum gravity or cosmological models should therefore be taken with a major
chunk of salt.

\subsection{Problem of time}
\label{s:Time}

Time in general relativity is observer-dependent and not absolute. Changing
time becomes part of the transformations in the hypersurface-deformation
algebroid as the fundamental symmetries of space-time. Accordingly, time
translations are not generated by an observable such as the energy, but by an
expression which on physical states always takes the value zero. As already
seen for the Friedmann equation in isotropic models, in terms of fundamental
fields and moments one obtains a constraint instead of an evolution
equation. The problem of time \cite{KucharTime,Isham:Time,AndersonTime} stems
from the fact that the evolution we naturally experience is hidden in the
formalism of a canonical theory and must be uncovered.

Deparameterization (Sec.~\ref{s:Deparam}) is a simple way of re-introducing
something that resembles evolution and time. However, the method is applicable
only in very special circumstances of specific matter contributions. Moreover,
there is a covariance problem in this context because different choices of
time variables are possible but difficult to compare. Usually, the resulting
quantum theories are so different that one cannot even find a unitary
transformation to formulate the question of how much their predictions vary
\cite{ReducedKasner,Electric,Search}. The most severe problem of
deparameterization, however, comes from the possibility of signature change in
consistent representations of the hypersurface-deformation algebroid: If one
deparameterizes, one merely picks a variable that classically depends
monotonically on time, and then views it as time itself. But quantization of
space-time can lead to the absence of time, which remains unseen if
deparameterization is used. 

In milder regimes, one can at least be sure that time exists. But for time to
be part of space-time, there must be strict transformations of covariance. In
particular, one must be able to change one's choice of time function without
affecting physical results. (Different observers must experience the same
physics.) If one uses deparameterization, one must therefore show that the
choice of time, such as $\phi$, does not matter. However, this property has
never been fully demonstrated in any cosmological (or other) model. Instead,
one can easily foresee obstructions to such a result: In order to formulate
deparameterized evolution, one manipulates the original constraint $C$,
usually involving a square root as in (\ref{apa}). These are simple
manipulations at the classical level, but different such versions require
non-trivial quantization choices. Since the quantum dynamics of cosmological
models can be rather complicated, one usually picks a quantization that looks
most simple. But a simple choice for one time may not relate to a simple
choice in another time. In this way, deparameterized models may break
covariance.

More realistic models are not deparameterizable by a global time. If one
solves for a momentum such as $p_{\phi}$, it typically is dependent on
``time'' $\phi$, unlike (\ref{apa}). The momentum is no longer preserved under
the evolution it generates, just like a time-dependent Hamiltonian. Turning
points of $\phi$ then become possible, where $p_{\phi}$ vanishes and the
classical $\phi$ starts running backwards. In a quantum model,
$\phi$-evolution then cannot be unitary, and is questionable even before the
turning point is reached.

Time coordinates are usually local. Similarly, one should expect any
$\phi$-time to be valid only for a finite range. For longer evolution, one
would have to find a way of transforming to a different time that remains
valid beyond the next turning point of $\phi$. Such transformations are
possible semiclassically \cite{EffTime,EffTimeLong,EffTimeCosmo}, although it
remains to be seen whether general quantum states enjoy the same feature.

As a shadow of non-unitary evolution, independence of physical results from
the choice of time requires time (as an expectation value of $\hat{\phi}$) to
be complex. One can see this feature, as well as general properties of the
effective constraint formalism, by solving some of the moment equations
implied by a constraint operator $\hat{C}=\hat{p}_{\phi}^2-
\hat{p}^2+V(\hat{\phi})$ for a relativistic particle in an arbitrary
$\phi$-dependent potential $V(\phi)$ \cite{EffTime}. If we try to use $\phi$
as time, we have to deal with a time-dependent Hamiltonian.

We compute the effective constraint $C_Q:=\langle\hat{C}\rangle$ up to second
order in moments:
\begin{equation}
 0=C_Q=\langle\hat{p}_{\phi}\rangle^2-\langle\hat{p}\rangle^2+
 \Delta (p_{\phi}^2)-\Delta (p^2)+V(\langle\hat{\phi}\rangle)+
 {\textstyle\frac{1}{2}}V''(\langle\hat{\phi}\rangle)\Delta (\phi^2)\,.
 \label{Cnon}
\end{equation}
However, this equation is not enough for a constrained system. The moments
are further restriced because $C_{\rm pol}=\langle\widehat{\rm
  pol}\hat{C}\rangle$ vanishes for any polynomial $\widehat{\rm pol}$ in basic
operators when an invariant state with $\hat{C}\psi=0$ is used.  We obtain
additional equations, including
\begin{equation}
0=C_{\phi}=
2\langle\hat{p}_{\phi}\rangle\Delta(\phi p_{\phi})+ i\hbar
\langle\hat{p}_{\phi}\rangle 
-2p\Delta(\phi p)+V'(\langle\hat{\phi}\rangle) \Delta (\phi^2) \label{Ct}\\
\end{equation}
The ordering of $\widehat{\rm pol}\hat{C}$ can, in general, not be chosen to
be symmetric because $\hat{C}$ must stay on the right for the product to
annihilate invariant states. Accordingly, we see some imaginary contributions
to the moment constraints.

Choosing a time variable $\phi$ by deparameterization means that it is no
longer treated as fluctuating, just like time in quantum mechanics. We can
implement this condition via moments by requiring $\Delta (\phi^2)=\Delta(\phi
q)=\Delta(\phi p)=0$ (a valid gauge fixing of the effective constrained
system).  Then, $\Delta(\phi p_{\phi})=-\frac{1}{2}i\hbar$ from
(\ref{Ct}). Other constraints of the type (\ref{Ct}) can be used to solve for
$\Delta(p_{\phi}^2)$ in (\ref{Cnon}), which latter acquires a non-trivial
imaginary part. For this to vanish, ${\rm Im}\,\langle\hat{\phi}\rangle=
-\frac{1}{2}\hbar/\langle\hat{p}_{\phi}\rangle\not=0$ to first order in
$\hbar$ \cite{EffTime,EffTimeLong}.

Covariance, in the presence of complex time, would then suggest that also
space is complex. While a covariant semiclassical picture of inhomogeneous
geometries remains to be worked out, there is a clear indication that complex
space-time models must be used. The physics of such systems is, however,
clear: In the formalism just described, observables, such as densities in
cosmology, are all real.

\subsection{Problem of states}
\label{s:States}

Cosmological observations do not tell us much about quantum states. Since
details of fundamental and effective descriptions often depend on the class of
states considered, they must be sufficiently general so as to be insensitive
to ambiguities related to undetermined states. This question plays a role a
two levels: for semiclassical as well as ground states.

\subsubsection{Semiclassical states.}

One's first idea of a semiclassical state is usually a sharply peaked
Gaussian. But such a state is very special: There is only one parameter (or at
most two if the state has correlations) which determines all infinitely many
moments. For a general semiclassical state, by contrast, any moment of order
$n$ is a free parameter as long as it is $O(\hbar^{n/2})$. Moreover, a sharply
peaked Gaussian is not a good late-time limit even in the case of a free
massive particle in quantum mechanics. Assuming a sharply-peaked Gaussian at
low curvature is therefore unjustified for several independent reasons.

The example of the free massive particle in quantum mechanics suggests a more
refined approach. Massive objects do not become more quantum if one just waits
long enough, for reasons which can be explained in a satisfactory way by
decoherence \cite{Decoherence}: The objects we usually deal with are not
completely isolated but are subject to weak interactions with a large number
of degrees of freedom, such as the molecules in surrounding air or the
photons, neutrinos and gravitational waves of cosmic backgrounds. When
these interactions, weak but acting over a long time, are integrated out, a
state different from the one of an isolated object is obtained. The resulting
state is no longer pure but mixed, and it approaches a near-diagonal density
matrix which can be interpreted in terms of classical probabilities. When
decoherence is taken into account, states need not be sharply-peaked Gaussians
in order to be in agreement with semiclassical behavior. In cosmology,
decoherence has been discussed in the context of inflationary structure
formation \cite{CosmoDecoh,QuantClassCosmo,PrimDecoh,PointerInflation} and for
quantum-cosmological questions \cite{FermionDecoherence}.

\subsubsection{Ground states.}

As seen for $C$ in isotropic models, Eq.~(\ref{Friedmann3}), the time
generator of gravity is unbounded from below. After quantization there is then
no ground state. The powerful low-energy method of perturbing around the
ground state is not available, and other distinguished states around which
moments would behave adiabatically are hard to find \cite{BouncePot}. In other
words, the derivative expansion of an effective action or Hamiltonian does not
exist under these circumstances. One must instead formulate effective systems
in which moments are kept as independent degrees of freedom, like the coupled
system (\ref{pdot}) with (\ref{dDq}), (\ref{dDqp}) and (\ref{dDp}). Such
systems can be analyzed and provide good information about the quantum
theory. But they cannot be brought to the form of an effective action with
higher-order time derivatives, or higher-curvature terms in gravity. Some of
the expectations of perturbative quantum gravity are therefore not realized in
general.

The absence of a ground state has important consequences for symmetries. An
effective theory is covariant under certain transformations only if the
zeroth-order state used to expand the quantum theory is invariant. (For
instance, a Poincar\'e-covariant effective action is obtained when one expands
around the invariant Minkowski vacuum.) It is, however, questionable whether
one can find exactly invariant states in a discrete space(-time) setting. Any
non-invariant choice would lead to an effective theory that is not covariant
under hypersurface deformations, and therefore is not generally covariant.

In order to find an invariant state, one would have to go to high excitation
levels of a many-patch theory, so that all possible patches are equally
excited. One would have to find some kind of ``finite-geometry'' state, in
analogy with finite-temperature states in quantum-field theory. Instead of the
temperature (or constituent velocities), geometrical quantities would have
non-zero expectation values of a uniform distribution. Candidates for such
states are being considered \cite{NoncyclicRep,NonDegVac}. The resulting
effective theories are likely to require quantum-corrected
hypersurface-deformation algebroids, as a reflection of the underlying
discreteness.

\section{Outlook: potential observations}

It is easier to break symmetries than to preserve them. In many examples of
condensed-matter physics, structures that violate the Euclidean
transformations of translations and rotations imply interesting effects which
can be used to test the theory or its effective descriptions, and even to
develop practical applications. The symmetries of space-time, however, are
more fragile when one considers quantum gravity. They can easily be broken too
(and have been broken many times in theoretical constructions). But if they
are violated, they do not give rise to physical effects but rather to an
inconsistent theory. Derivations of potentially observable effects in quantum
gravity must therefore proceed much more carefully.

\subsection{The problem of the Planck scale}

The Planck scale is extreme and cannot be reached by any experiments in the
foreseeable future. Dimensional arguments that only consider the magnitude of
this scale in relation to typical observables are often taken to indicate that
quantum gravity is unobservable. Moreover, since dimensional arguments work
only if there is a small number of relevant parameters with the same
dimension, such as the fundamentally defined Planck energy and the contingent
average density of the universe, different types of corrections are often
estimated to be of the same size. A careful analysis of their interplay is
then required to derive reliable consequences.

In the Planck regime, however, any such detailed analysis suffers from the
ambiguity problem of quantum gravity. Even if one restricts attention to just
one proposal of a general theory, several quantization and other choices still
have to be made for a concrete model of the early universe. Most of the time,
there is not much more than one's preference or prejudice that selects crucial
ingredients. (See for instance the discussion of different times in
deparameterized models, Sec.~\ref{s:Time}.) The form of a single quantum
correction, and even more so the combined effects of different corrections,
can depend sensitively on such choices, and widely varying scenarios can be
obtained from the same theory, making the Planck regime highly
uncontrolled. Since no direct observations exist to limit the choices, with
such models it is hard to improve our understanding of quantum gravity.  This
ambiguity problem prevents reliable physics to be drawn from direct
considerations of the Planck era, where quantum-gravity corrections would be
large.

For reliable input, one must look at tamer regimes with at least some indirect
observational access. Such regimes have small density, compared to the Planck
density, and direct quantum-gravity effects are expected to be tiny. However,
the description of such regimes is more involved than the comparison of two
parameters as used in dimensional arguments. At lower densities, the universe
has grown out of its Planckian compactness, and developed some structure as
envisioned for instance in the inflationary scenario. Not just the average
density but also parameters that describe finer details are then
relevant. They should only have a weak influence on quantum-gravity effects,
but they can act for a long, cosmic time and eventually make themselves
noticeable. Such features would have to be derived from the theory and could
not be guessed by dimensional arguments.

Dimensional arguments typically fail in realistic settings. For instance,
while it is easy to find the (Bohr) radius of a hydrogen atom based only on
the relevant constants of nature, the radius of heavier atoms, with many
constituents in their nuclei and shells, can only be computed from a detailed
theory. Similarly, quantum gravity may well give rise to effects that violate
expectations based on dimensional arguments. Discrete theories, for instance,
have a large number of spatial or space-time constituents which make
dimensional arguments fail just like heavy atoms do. In addition to the Planck
density and the average matter density, for instance, a third relevant
parameter could be the density of discrete patches that make up an expanding
spatial region. 

In order to show convincingly that additional parameters not only render
dimensional arguments inapplicable but also lead to more-sizeable effects, one
must find concrete mechanisms by which magnifications can occur, perhaps by
the concerted action of many constituents. There is at least one indication
from quantum-corrected algebroids of hypersurface-deformation: By taking into
account discrete or other features, they provide non-classical space-time
structures and therefore lead to effective theories more general than standard
gravitational effective actions with their tiny higher-curvature terms.

\subsection{Indirect effects}

We just list here some candidates that have been considered for possible
enlargements of tiny quantum effects. 

The most well-known example is perhaps large extra dimensions in which only
gravity but not the other forces can penetrate \cite{Hierarchy}. The latter
requirement ensures that such models are not in conflict with observations in
particle physics. The former, of large extra dimensions, provides a parameter
that could lift corrections to observable magnitudes.

Another example is new fields that may serve as candidates for an inflaton.
(See for instance \cite{NonSingdeSitter,RelaxLambda}.)
In addition to the advantage of having an explanation of inflation, such
models can make their underlying theories testable by using the precise
measurements of the cosmic microwave background.

Finally, discrete theories provide a clear candidate for constituents of large
number. If their independent impacts on space-time physics, for instance of a
propagating wave, add up, the result may be observable. Such an effect would
be analogous to Brownian motion, which made it possible to prove the existence
of atoms and molecules without having atomic resolution. In quantum gravity,
one needs long distance or time scales to have a large number of constituents,
which can be realized in signals from gamma-ray bursts or by structure
building up slowly as the early universe expands.
Some reviews of concrete realizations can be found in
\cite{DeformedQG,ObsLQC}.

\section*{Summary}

Space-time presents a challenging object for quantum and effective
descriptions, posing interesting questions. We could here discuss only a small
part of the large body of work done on quantum cosmology, focusing instead on
general problems, principles, and procedures. In most cases of discussions in
the literature, at least one of the relevant requirements on quantum cosmology
is violated, indicating how difficult progress in this field still is.
For instance, the problem of time is circumvented in approaches that make use
of a background metric, while in background independent ones it is often
evaded by choosing one internal time but not asking whether results are
independent of the choice. The problem of states is similarly avoided (but not
solved) by assuming specific states, such as Gaussians, or by using low-energy
expansions around vacuum states of free theories even in strong quantum
regimes. The covariance problem is solved in some approaches, such as string
theory, but presents one of the most crucial and stubborn issues for canonical
and discrete space-time formulations.

We remain far from a proper understanding of quantum cosmology, especially
when physics at the Planck scale is involved. At the same time, research on
quantum cosmology has led to progress in our understanding of generally
covariant quantum systems and often showed unexpected effects of quantum
space-time.

\section*{Acknowledgements}

This work was supported in part by NSF grant PHY-1307408. The author is
grateful to Beverly Berger for the invitation to write this review.


\end{document}